\documentclass[prl,reprint,amsmath,amssymb,aps,superscriptaddress]{revtex4-2}

\usepackage[T1]{fontenc}
\usepackage{lmodern}
\usepackage{graphicx}
\usepackage{bm}
\usepackage{xcolor}

\newcommand{\gsp}{g_{\mathrm{sp}}}
\newcommand{\meff}{m_{\mathrm{eff}}}
\newcommand{\PR}{\mathrm{PR}}

\begin{document}

\title{In-Band Boundary Resonances from Coalescing Zero-Group-Velocity Points}

\author{Amir Rajabpoor Alisepahi}
\affiliation{Department of Mechanical Engineering,
  University of Vermont, Burlington, Vermont 05405, USA}

\author{Charles A.\ Downing}
\affiliation{Department of Physics and Astronomy,
  University of Exeter, Exeter EX4 4QL, United Kingdom}

\author{Jihong A. Ma}
\email{Jihong.Ma@uvm.edu}
\affiliation{Department of Mechanical Engineering,
  University of Vermont, Burlington, Vermont 05405, USA}
\affiliation{Department of Physics,
  University of Vermont, Burlington, Vermont 05405, USA}
\affiliation{Materials Science Program,
  University of Vermont, Burlington, Vermont 05405, USA}

\begin{abstract}
Boundary states generally reside in band gaps, whereas confinement within a propagating band typically requires exact decoupling from the continuum, as in bound states in the continuum. Here we demonstrate bulk-coupled, boundary-weighted resonances within the bands of reciprocal, topologically trivial, Hermitian lattices. They arise from coalescing zero-group-velocity points: the merged Bloch root remains propagating but, at nearby energies, splits into weakly evanescent channels. The parity of its multiplicity determines the boundary response. At a quartic coalescence, the merged root yields only evanescent channels on one side of the coalescence, enabling a bulk-matched termination to form a resonance whose evanescent weight grows as the cube of the decay length, rather than linearly as at an isolated zero-group-velocity point. At a cubic coalescence, by contrast, each evanescent channel locks to a propagating partner, depleting the boundary despite a diverging density of states. We realize quartic coalescence in minimal and longer-range ladders, distinguish the resonance from extended and defect-bound states through finite-size scaling, and observe boundary enhancement just below the nearly flat zone-boundary frequency using scanning laser Doppler vibrometry on a 3D-printed elastic ladder. Bloch-root multiplicity thus provides a platform-independent design principle for in-band boundary resonances.
\end{abstract}

\maketitle

Wave propagation and localization in periodic structures govern the transport of electrons, lattice vibrations, and electromagnetic signals. Ordinarily, a wave can remain near a boundary only when no propagating bulk wave is available to carry it away, as in a spectral gap. Modified terminations, band inversion, and local defects produce Tamm, Shockley, and defect states in such gaps~\cite{tamm1932,shockley1939,davison1992}, and nontrivial bulk invariants guarantee topological edge states~\cite{kane2005,huber2016,xue2022,ma2019,rajabpoor2023breakdown,rajabpoor2024gap,pyfrom2025hierarchical,ozawa2019,piwowarczyk2025} or, in Maxwell lattices, floppy modes at zero frequency~\cite{kane2014,ma2018}. Inside a propagating band, a boundary excitation instead leaks into bulk waves at the same frequency, leaving at most a surface resonance~\cite{davison1992}. Bound states in the continuum (BICs)~\cite{vonNeumann1929} avoid leakage by decoupling exactly from all propagating waves through symmetry, interference, or Fabry--P\'{e}rot conditions~\cite{hsu2016,sadreev2021,li2019,sadrieva2019,fano1961,miroshnichenko2010,gao2023,liu2023,yan2025,haq2021,rahman2022,benalcazar2020,cerjan2020}, and become quasi-BICs with finite lifetimes when this decoupling is broken~\cite{koshelev2019,koshelev2023}; compact localized states in flat bands are likewise decoupled~\cite{leykam2018}. These mechanisms leave open a basic question: can a resonance remain coupled to propagating waves yet concentrate at an otherwise unmodified boundary, without a gap, an edge potential, exact decoupling, or the non-Hermiticity behind the skin effect~\cite{yao2018} (Table~\ref{tab:compare})? Such a state would be excited by propagating bulk waves at the same frequency yet concentrate their intensity at the boundary, a combination that otherwise requires an edge potential: gap states have no propagating bulk waves to couple to, and an ideal BIC is, by construction, invisible to them.

\begin{table}[t]
\caption{\label{tab:compare}Boundary-localization mechanisms in Hermitian lattices. PR is the participation ratio of an $N$-cell lattice, and $W_b$ the fraction of modal weight in the $n_b$ cells nearest the two ends, with uniform value $2n_b/N$~\cite{sm}.}
\footnotesize
\begin{ruledtabular}
\begin{tabular}{p{1.9cm}p{0.6cm}p{1.3cm}p{2.0cm}p{1.7cm}}
State & Gap & Continuum coupling & Localization & Origin \tabularnewline
\hline
Gap states (topological, Tamm, defect)~\cite{kane2005,tamm1932,shockley1939} & Yes & None & Exponential, $\PR\propto N^{-1}$ & Bulk invariant or edge potential \tabularnewline
Surface resonance~\cite{davison1992} & No & Yes & Edge weighted, finite lifetime & Edge potential \tabularnewline
BIC, quasi-BIC~\cite{hsu2016,koshelev2019} & No & None or weak & Bound, or finite lifetime & Symmetry or interference \tabularnewline
This work & No & Yes & $W_b\gg2n_b/N$; $\PR$ falls slower than $N^{-1}$ & Quartic Bloch root, matched edge \tabularnewline
\end{tabular}
\end{ruledtabular}
\end{table}

Boundary-weighted modes of this kind were recently found numerically near flat portions of the bands of a two-leg bosonic ladder~\cite{downing2024}, but their origin, and whether they were bound states or resonances, remained unclear. A clue comes from zero-group-velocity (ZGV) points, the stationary points of a band. Isolated ZGVs support local and edge resonances in elastic plates~\cite{prada2008local,ces2011edge} and raise quality factors when combined with BICs~\cite{zgvbic2024}. Merging ZGVs produce cubic and quartic dispersions, which underlie frozen light, slow light, gain enhancement, and higher-order transfer-matrix degeneracies~\cite{figotin2006frozen,figotin2006slow,othman2016giant,othman2017epd,nada2017crow}. These studies concern bulk propagation or transmission, and their finite-structure resonances are extended standing waves. A conventional degenerate band edge (DBE) is also unsuitable: its quartic point terminates the propagating branch rather than lying within a band. Whether a band singularity can concentrate a continuum-coupled resonance at a termination, and which singularities can, has remained open; flatness alone does not decide it.

Here we show that the deciding property is the multiplicity of the coalesced Bloch root. A Bloch root describes how a wave changes from one unit cell to the next~\cite{brillouin1946}: roots of unit magnitude propagate, and all others decay or grow. Near a coalescence, the merged root splits into a cluster of propagating or weakly evanescent channels, and the parity of its multiplicity determines whether the boundary is enhanced or depleted. At a quartic coalescence, the first and most accessible even case beyond the double root of an isolated ZGV, the cluster contains only evanescent channels on one side of the coalescence frequency, and a bulk-matched termination combines them into a boundary-weighted resonance. We realize quartic points at the zone boundary and in the band interior, and observe the predicted enhancement in a 3D-printed elastic ladder.

For a one-dimensional lattice with Bloch Hamiltonian $\mathcal{H}(\lambda)=\sum_rH_r\lambda^r$, where the Bloch factor $\lambda$ relates neighboring unit cells of length $a$ through $\psi_{n+1}=\lambda\psi_n$, the Bloch roots $\lambda_j(E)$ satisfy
\begin{equation}
	P(E,\lambda)\equiv\det[\mathcal{H}(\lambda)-EI]=0.
	\label{eq:char}
\end{equation}
Roots on the unit circle, $\lambda=e^{iqa}$, describe propagation, whereas roots away from it describe evanescence. At a regular band point ($\partial_EP\neq0$), implicit differentiation of $P[E(q),e^{iqa}]\equiv0$ gives~\cite{sm}
\begin{equation}
	\frac{d^jE}{dq^j}(q_0)=0\;\;(j\le k)
	\;\Longleftrightarrow\;
	\partial_\lambda^jP(E_0,\lambda_0)=0\;\;(j\le k).
	\label{eq:cond}
\end{equation}
A ZGV ($k=1$) is thus a repeated Bloch root on the unit circle. A coalescence of two ZGVs ($k=2$), at which the inverse effective mass $1/\meff\equiv d^2\omega/dq^2$ also vanishes, raises the root multiplicity $m$ to at least three, and $k=3$ raises it to at least four.

For a reciprocal two-leg ladder with intrachain coupling ranges $r_A$ and $r_B$, time-reversal symmetry pairs $\lambda$ with $\lambda^{-1}$, so $P(E,\lambda)=Q(E,c)$ with $c=\cos qa=(\lambda+\lambda^{-1})/2$, where $Q$ has degree $M=r_A+r_B$ in $c$; $M$ is also the number of boundary conditions at a termination, one per site beyond the cut~\cite{sm}. Each root $c$ defines a channel: a real root in $[-1,1]$ gives a counterpropagating pair $\lambda=e^{\pm iqa}$, and any other root gives one factor with $|\lambda|<1$, decaying from a left boundary with decay length $\xi=-a/\ln|\lambda|$. The number of evanescent channels is $d=M-N_{\mathrm{prop}}$, where $N_{\mathrm{prop}}$ counts the real roots in $[-1,1]$.

The two lowest-codimension coalescence mechanisms are relevant in the reciprocal one-dimensional setting considered here. At an interior wave number $q^*$, a neighboring maximum and minimum can merge, making each of the time-reversed Bloch factors $\lambda=e^{\pm iq^*a}$ a triple root, with $E-E^*\propto(q-q^*)^3$. At $q=0$ or $\pi/a$, a ZGV can instead merge with its time-reversed partner at the preexisting extremum. Symmetry forces the odd dispersion derivatives to vanish at these wave numbers, allowing a quartic root; at $q=\pi/a$, for example, $E-E_\pi\propto(q-\pi/a)^4$, and $\lambda=-1$ is a fourfold root. Both mechanisms have codimension one, occurring at isolated values along a generic one-parameter sweep, whereas a quartic coalescence at a generic interior wave number has codimension two and occurs at isolated points in a two-parameter plane. More generally, an interior $m$-fold root has codimension $m-2$, whereas a root at a time-reversal-invariant wave number must have even multiplicity and has codimension $(m-2)/2$~\cite{sm}. The zone-boundary quartic point is thus the most accessible nontrivial even-multiplicity coalescence.

At nearby energies, the local relation $A\,\delta E+D\,\delta q^m=0$, with real $A$ and $D$, splits it into $m$ roots, $\delta q=(-A\,\delta E/D)^{1/m}e^{2\pi ik/m}$, separated in $qa$ by $O(a/\xi)$, with $\xi\propto|\delta E|^{-1/m}$. A termination acting over a few cells therefore couples them collectively. For odd $m$, the cluster always contains a propagating root: at a cubic point, boundary matching locks the decaying amplitude to cancel its propagating partner, $\psi_n\simeq\lambda_p^n-\lambda_d^n\propto na/\xi$, and the boundary is depleted for any generic local termination, despite a diverging density of states~\cite{sm}. For even $m$, one side of $E_0$ has $m/2$ decaying roots and no propagating one. At an isolated ZGV ($m=2$), the single decaying factor gives an evanescent weight $W$, measured in units of the bulk intensity per cell, that scales as $\xi$; at a quartic point, the two decaying factors combine into $\lambda_0^ne^{-na/\xi}$ and $n\lambda_0^{n-1}e^{-na/\xi}$, giving $W\propto\xi^3$~\cite{sm}. 

\begin{figure}[t]
	\includegraphics[width=\columnwidth]{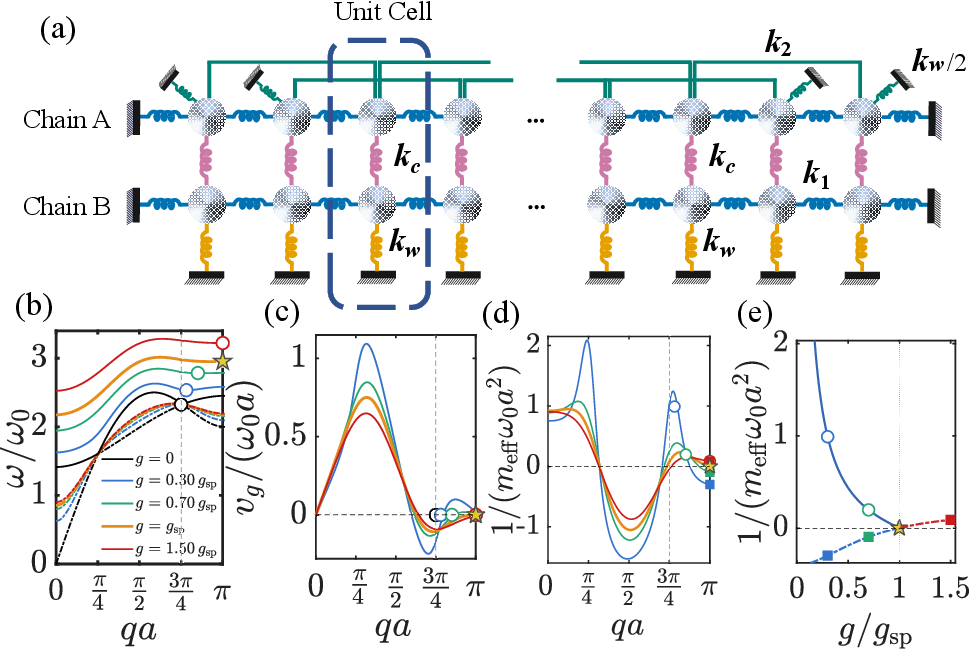}
	\caption{\label{fig:model}Minimal ladder and zone-boundary self-pairing. (a) Unit cell. (b) Bands $\omega_\pm/\omega_0$ (upper solid, lower dashed) at $\eta=1$ for $g/\gsp=0$, 0.3, 0.7, 1.0, and 1.5; circles mark the interior ZGV of the upper band, which reaches $qa=\pi$ at $\gsp$ (star). (c)--(e) Group velocity and inverse effective mass at the interior ZGV (circles) and zone boundary (squares); vertical lines, bare-band crossing.}
\end{figure}
\begin{figure*}[t]
	\includegraphics[width=\textwidth]{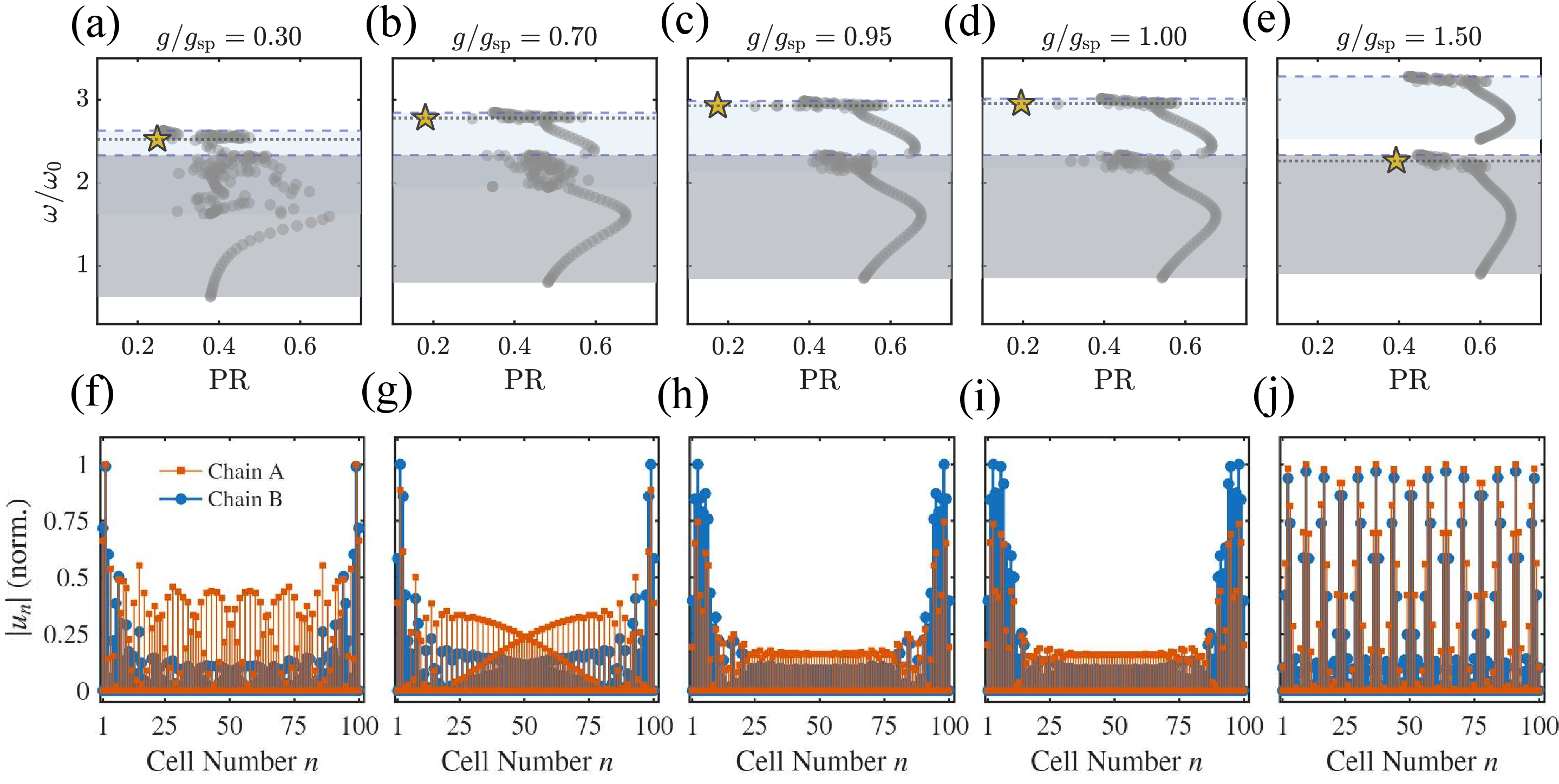}
	\caption{\label{fig:spectra}Finite-size spectra across self-pairing ($N=100$, $\eta=1$). (a)--(e) PR versus $\omega/\omega_0$ for $g/\gsp=0.3$, 0.7, 0.95, 1.0, and 1.5; gray and blue shading mark the lower and upper bands, and stars the minimum-PR modes, whose amplitudes $|u_{nA}|$ (orange) and $|u_{nB}|$ (blue) are shown in (f)--(j).}
\end{figure*}

\begin{figure}[t]
	\includegraphics[width=\columnwidth]{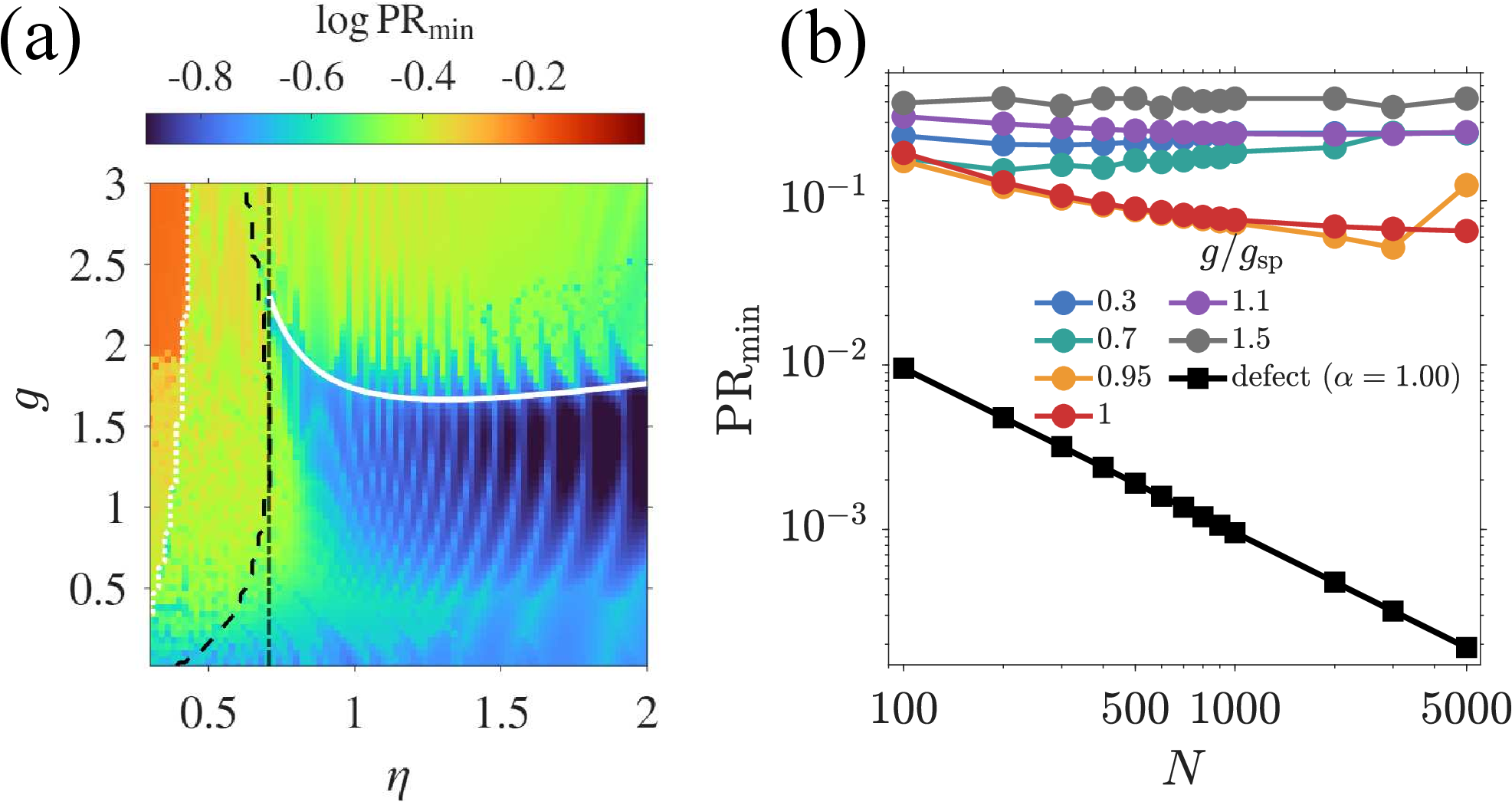}
	\caption{\label{fig:loc}(a) $\log_{10}\PR_{\min}$ over $(\eta,g)$ for $N=100$. White solid curve, $\gsp(\eta)$; dashed and dotted curves, onsets of additional ZGVs; dash-dotted line, $\eta=1/\sqrt2$. (b) $\PR_{\min}$ versus $N$ at $\eta=1$ (upturn at the largest $N$: weight sharing~\cite{sm}). Black: terminal-defect bound state ($k_{\mathrm{def}}=3k_1$), with $\PR_{\min}\propto N^{-1.00}$.}
\end{figure}
\begin{figure*}[t] \includegraphics[width=\textwidth]{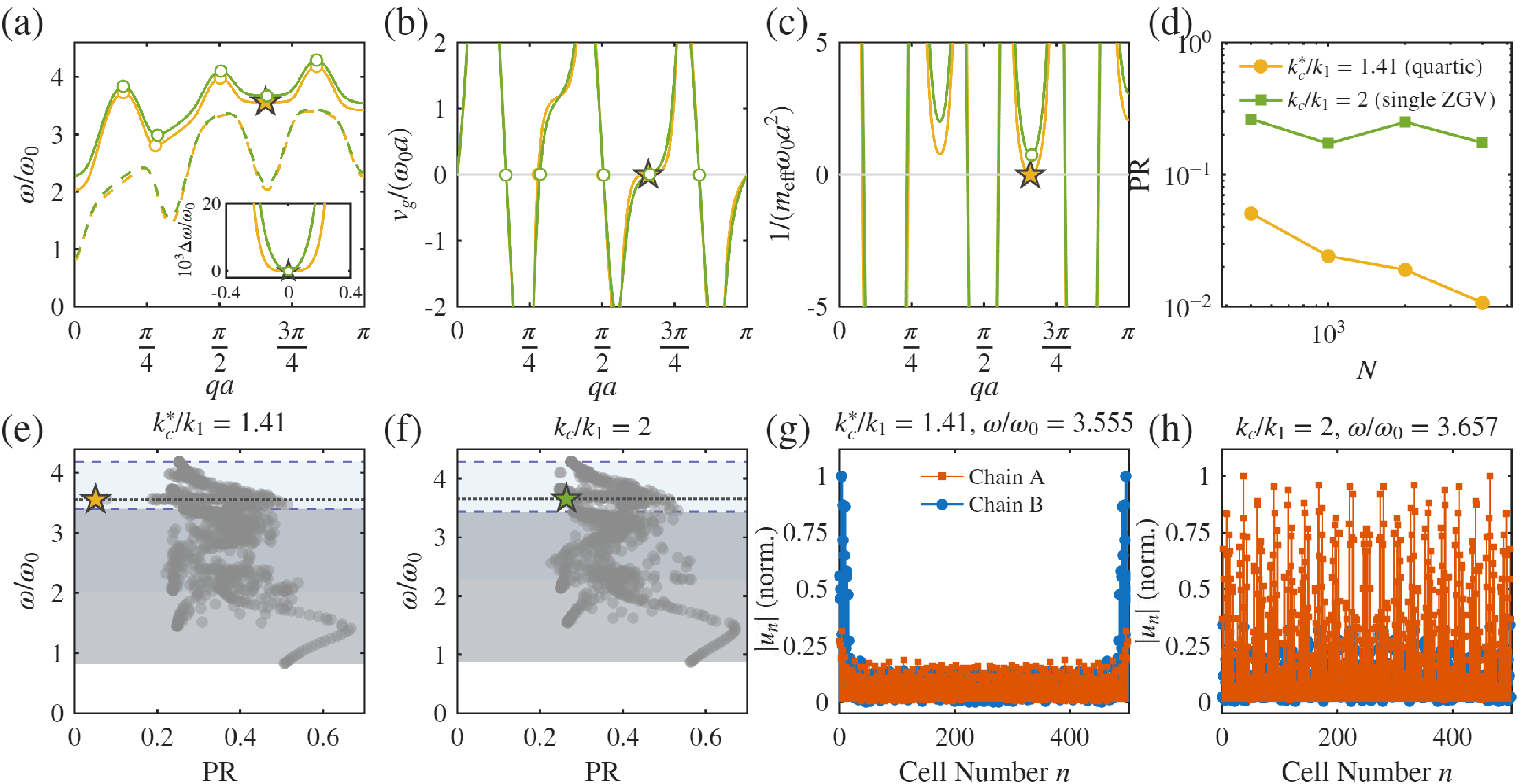} \caption{\label{fig:interior}Interior quartic point ($k_c^*/k_1=1.41$, $k_{2B}^*/k_1=1.014$; yellow) versus a single ZGV ($k_c/k_1=2$; green). (a) Bands, (b) group velocity, and (c) inverse effective mass; circles mark ZGVs, the star their merger. (d) Minimum PR just below $\omega_c$ versus $N$. (e),(f) PR spectra and (g),(h) starred-mode amplitudes for $N=500$.}
\end{figure*}
\begin{figure*}[t]
	\includegraphics[width=\textwidth]{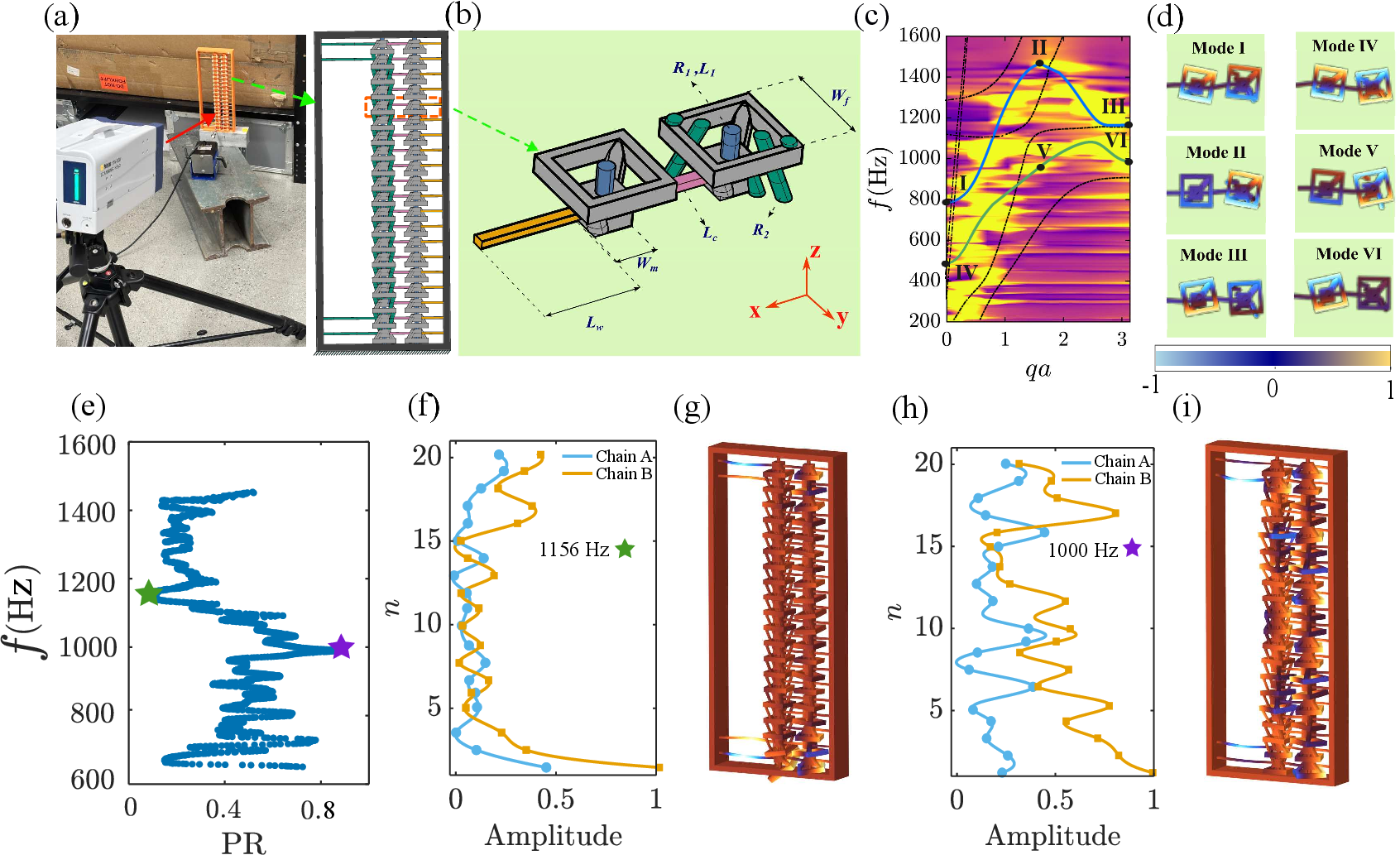}
	\caption{\label{fig:exp}Experiment. (a) PLA ladder in the SLDV setup and (b) unit cell. (c) Measured dispersion (color) and computed torsional branches (solid; dashed, other modes); (d) mode shapes at points I--VI. (e) Simulated PR for $N=20$; stars mark 1156 and 1000~Hz. (f),(h) Measured velocity amplitudes at these frequencies ($n=1$ is the actuated cell) and (g),(i) simulated modes.}
\end{figure*}
We realize this mechanism in the spring--mass ladder of Fig.~\ref{fig:model}(a), with nearest-neighbor (NN) stiffness $k_1$ on both chains, next-nearest-neighbor (NNN) stiffness $k_2$ on chain A, rung stiffness $k_c$, and grounding $k_w=2k_2$ on chain B, which equalizes the bulk diagonal stiffnesses, making it a phononic analogue of the bosonic ladder of Ref.~\cite{downing2024}. With $\eta=k_2/k_1$, $g=k_c/k_1$, and $\omega_0^2=k_1/m$, the eigenproblem for $E=\omega^2$ is that of a two-leg tight-binding Hamiltonian, so the results apply to any lattice with these couplings~\cite{sm}. Here $M=3$; the model is reciprocal and Hermitian and belongs to class AI, which has no nontrivial stable invariant in one dimension.

Rung coupling hybridizes the counterpropagating bare bands and creates an interior ZGV in the upper band, which for $\eta>1/\sqrt2$ moves toward $qa=\pi$ as $g$ increases [Fig.~\ref{fig:model}(b)--(e)]~\cite{sm}. Applying Eq.~\eqref{eq:cond} with $k=2$ yields interior cubic coalescences for $1/(2\sqrt2)<\eta\le1/\sqrt2$, which deplete the boundary~\cite[Fig.~S1]{sm}, and zone-boundary self-pairing at
\begin{equation}
	\gsp(\eta)=\sqrt{\left(\frac{2\eta^2}{2\eta-1}\right)^2-\eta^2},\qquad\eta>\frac{1}{\sqrt2}.
	\label{eq:gsp}
\end{equation}
At $\eta=1$, $\gsp=\sqrt3$. For $g<\gsp$, the zone boundary is a weak maximum flanked by interior minima at $\pm q_s$; at $g=\gsp$, these three ZGVs merge into a quartic root; for $g>\gsp$, the zone boundary becomes an ordinary minimum. Throughout, an interior maximum near $qa\simeq1.9$ keeps $\omega_+(\pi/a)$ inside the upper band, so, unlike a conventional DBE, the quartic point is embedded in the continuum.

We terminate an $N$-cell ladder by cutting the intercell bonds while retaining the bulk diagonal terms. This bulk-matched termination is the standard open tight-binding boundary and adds no edge potential; in the phononic lattice, grounding springs restore the diagonal stiffness lost at the cut~\cite{sm}. We characterize each eigenmode by its participation ratio PR, which is unity for a uniform mode, and by its boundary weight $W_b$ (Table~\ref{tab:compare})~\cite{sm}; $\PR_{\min}$ denotes the minimum PR over all modes.

Figure~\ref{fig:spectra} follows the spectrum across self-pairing at $\eta=1$. Below $\gsp$, and again well above it in the lower band, the minimum-PR mode is extended, with a pronounced bulk envelope [Fig.~\ref{fig:spectra}(f),(g),(j)]. Near $\gsp$, the bulk envelope collapses to a weak plateau beneath strong boundary peaks [Fig.~\ref{fig:spectra}(h),(i)], with $W_b=0.83$ at $\gsp$. Across the $(\eta,g)$ plane, low $\PR_{\min}$ occurs for $\eta>1/\sqrt2$ and is bounded above by $\gsp(\eta)$ [Fig.~\ref{fig:loc}(a)]~\cite{sm}.

Finite-size scaling separates this mode from extended and bound states [Fig.~\ref{fig:loc}(b)]. Away from self-pairing, $\PR_{\min}$ stays between 0.15 and 0.45 without decreasing with $N$, as for extended states, whereas a terminal-defect bound state follows $\PR_{\min}\propto N^{-1}$. At $\gsp$, $\PR_{\min}$ instead falls from 0.20 at $N=100$ to 0.07 at $N=4000$ in a slow crossover, while $W_b=0.20$ at $N=4000$, 40 times its uniform value~\cite{sm}. The mode thus occupies a growing number of cells yet stays strongly boundary weighted: it is a continuum resonance, not a bound state. These sizes remain in a crossover without a simple PR power law; the asymptotic $\xi^3$ law concerns $W$, not PR~\cite{sm}.

The Bloch channels explain this behavior. At the $N=100$ resonance [Fig.~\ref{fig:spectra}(d),(i)], just below $\omega_\pi\equiv\omega_+(\pi/a)=2.955\,\omega_0$, the quartic root has split into two weakly evanescent channels with $\xi/a\simeq5.8$, both descending from $\lambda=-1$, while the remaining channel propagates at $qa\simeq\pi/2$ and forms the bulk plateau~\cite{sm}. Far from this propagating channel, the evanescent channels are not locked, unlike at a cubic coalescence~\cite[Fig.~S1]{sm}. Because the $d=2$ decaying channels cannot generically satisfy the $M=3$ boundary conditions while a propagating channel remains, the termination yields a resonance rather than a bound state. The termination sets the strength of the resonance, not the $\xi^3$ growth of $W$~\cite{sm}.

The mechanism requires neither the zone boundary nor self-pairing. We show this in an asymmetric ladder with sixth-neighbor coupling on chain A and NNN coupling $k_{2B}$ on chain B, giving $M=8$~\cite{sm}. Tuning two stiffnesses, as the codimension-two count requires, to $k_c^*/k_1=1.408$ and $k_{2B}^*/k_1=1.014$ merges two upper-band minima and the maximum between them at $q^*a=2.07$ and $\omega^*=3.558\,\omega_0$, where $\cos q^*a$ becomes a quadruple root of $Q$ and $d^3\omega/dq^3$ also vanishes [Fig.~\ref{fig:interior}(a)--(c)].

For $N=500$, the minimum-PR mode lies just below $\omega^*$, with $\PR=0.051$ and $W_b=0.66$ [Fig.~\ref{fig:interior}(e),(g)]. Its four decaying factors, two from each time-reversed cluster, descend from the quartic root ($\xi/a\simeq9$), while the four propagating channels lie elsewhere in the zone, so the evanescent channels are not locked and the mode is again a resonance~\cite{sm}: PR falls more slowly than $N^{-1}$, to 0.011 at $N=4000$, as $2N\PR$ grows to 88 sites and $W_b$ to 33 times its uniform value [Fig.~\ref{fig:interior}(d)]. At $k_c=2k_1$, where only an ordinary minimum remains near $q^*$, the mode stays extended, with $\PR\simeq0.2$ [Fig.~\ref{fig:interior}(f),(h)]. At a cubic coalescence of two of these ZGVs, by contrast, the boundary weight is carried by an adjacent isolated minimum, and PR levels off near 0.04, nearly four times the quartic value at $N=4000$~\cite[Fig.~S2]{sm}.


To test the mechanism experimentally, we 3D-printed a 20-cell elastic realization of the minimal ladder with an approximately bulk-matched termination [Fig.~\ref{fig:exp}(a),(b)], drove one end, and measured the torsional motion of both chains by scanning laser Doppler vibrometry (SLDV)~\cite{sm}. The measured and simulated torsional dispersions reproduce the predicted band shapes, including a nearly flat zone boundary at $f_\pi=1181.5$~Hz [Fig.~\ref{fig:exp}(c),(d)], which places the specimen near $g/\gsp\simeq0.9$, where the weakly evanescent channels persist~\cite{sm}. The simulated PR is minimized at 1156~Hz, just below $f_\pi$ [Fig.~\ref{fig:exp}(e)]. There the measured response is concentrated at the driven boundary: excluding the actuated cell, the two adjacent cells hold a fraction $W_b^{\mathrm{low}}=0.59$ of the measured weight, a one-sided measure distinct from $W_b$ [Fig.~\ref{fig:exp}(f),(g)], against 0.25 at 1000~Hz, where the response extends throughout the ladder [Fig.~\ref{fig:exp}(h),(i)], and 0.11 for a uniform response~\cite{sm}. Because both responses are driven at the same cell, proximity to the actuator cannot account for the 2.4-fold difference. The experiment thus reveals the frequency-selective boundary enhancement predicted on the near-critical, weakly evanescent side of the quartic coalescence.

ZGV coalescence is therefore a bulk organizing center but not a sufficient condition: the parity of the coalesced root's multiplicity decides whether its evanescent channels can concentrate at a boundary, and the termination sets how strongly. The resulting design rule for one-dimensional reciprocal Hermitian lattices is to operate near an even-multiplicity coalescence, most accessibly a quartic point at a time-reversal-invariant wave number, and to minimize edge detuning, ideally with a bulk-matched termination. The criterion should extend to higher dimensions at fixed wave number parallel to the boundary; given mirror symmetry normal to the boundary, this conserved wave number can supply the control parameter for a symmetry-assisted quartic point. Candidate platforms include waveguide arrays~\cite{szameit2009zigzag}, polaritonic lattices~\cite{amo2016polariton}, electrical circuits~\cite{carusotto2020circuit,lee2018topolectrical}, atomic lattices~\cite{bloch2008many}, and elastic plates near ZGV Lamb modes~\cite{prada2008local,ces2011edge}. Even-multiplicity Bloch-root coalescences thus offer a platform-independent route to controlling in-band boundary responses.

\begin{acknowledgments}
	ARA and JAM acknowledge the support of National Science Foundation Award No. 2119485. CAD is supported by the Royal Society via a University Research Fellowship (URF/R1/201158). JAM is also partially supported by National Aeronautics and Space Administration (NASA Grant 80NSSC23M0071) and National Science Foundation CAREER Award 2542612.
\end{acknowledgments}

\bibliography{references}

\clearpage
\onecolumngrid
\begin{center}
{\large\bfseries Supplemental Material:\\[2pt]
In-Band Boundary Resonances from Coalescing Zero-Group-Velocity Points}\\[8pt]
Amir Rajabpoor Alisepahi,$^{1}$ Charles A.\ Downing,$^{2}$ and Jihong A. Ma$^{1,3,4}$\\[4pt]
{\small\itshape
$^{1}$Department of Mechanical Engineering, University of Vermont, Burlington, Vermont 05405, USA\\
$^{2}$Department of Physics and Astronomy, University of Exeter, Exeter EX4 4QL, United Kingdom\\
$^{3}$Department of Physics, University of Vermont, Burlington, Vermont 05405, USA\\
$^{4}$Materials Science Program, University of Vermont, Burlington, Vermont 05405, USA}
\end{center}
\vspace{6pt}
\twocolumngrid
\setcounter{secnumdepth}{1}
\setcounter{section}{0}\setcounter{equation}{0}\setcounter{figure}{0}\setcounter{table}{0}
\renewcommand{\thesection}{S\arabic{section}}
\renewcommand{\theequation}{S\arabic{equation}}
\renewcommand{\thefigure}{S\arabic{figure}}
\renewcommand{\thetable}{S\arabic{table}}

\section{Dispersion derivatives and root multiplicity}
\label{sec:mult}

We write the Bloch solution as $\psi_n=\lambda^n\mathbf{e}$, where $\psi_n$ collects the amplitudes of the $\nu$ sites in cell $n$ ($\nu=2$ for the ladders considered here) and $\mathbf{e}$ is the polarization vector, so that $P(E,\lambda)\equiv\det[\mathcal{H}(\lambda)-EI]=0$ [main-text Eq.~(1)]. For Hermitian $H_r$ and real $E$, $P(E,e^{iqa})$ is real. Differentiating $P[E(q),e^{iqa}]\equiv0$ with respect to $q$, with $d\lambda/dq=ia\lambda$, gives
\begin{equation}
	\partial_EP\,\frac{dE}{dq}+ia\lambda\,\partial_\lambda P=0.
	\label{eq:d1}
\end{equation}
Away from band touchings $\partial_EP\neq0$, and $\lambda\neq0$, so $dE/dq=0$ if and only if $\partial_\lambda P=0$, which is main-text Eq.~(2) with $k=1$. Differentiating Eq.~\eqref{eq:d1} once more and evaluating at a point where $dE/dq=\partial_\lambda P=0$, all cross terms vanish:
\begin{equation}
	\partial_EP\,\frac{d^2E}{dq^2}+(ia\lambda_0)^2\,\partial_\lambda^2P=0,
	\label{eq:d2}
\end{equation}
so that, at a ZGV, $d^2E/dq^2=0$ if and only if $\partial_\lambda^2P=0$, which is main-text Eq.~(2) with $k=2$. Iterating, $d^jE/dq^j=0$ for $j=1,\dots,k$ is equivalent to $\partial_\lambda^jP=0$ for $j=1,\dots,k$, which is main-text Eq.~(2) for general $k$. At $q=\pi/a$, where $E(q)$ is even, all odd derivatives vanish identically, so the $k=3$ condition adds no constraint there.

At fixed $E_0$, these are derivatives of the polynomial $p(\lambda)=P(E_0,\lambda)$. A double root, $p=(\lambda-\lambda_0)^2R(\lambda)$ with $R(\lambda_0)\neq0$, has $p''(\lambda_0)=2R(\lambda_0)\neq0$, so main-text Eq.~(2) with $k=2$ forces $\lambda_0$ to be a root of multiplicity at least three; by the same argument, $k=3$ forces multiplicity at least four. The multiplicity is generically three at an interior (saddle-node) coalescence of two ZGVs and four at a coalescence of three ZGVs, which has codimension one at the zone boundary and two in the interior (see below and Sec.~\ref{sec:roots}). Its parity decides whether the nearby evanescent channels can form a boundary resonance (Sec.~\ref{sec:locking}).

\textit{Codimension.}---The codimension of a coalescence is the number of independent conditions it imposes on the lattice parameters. A codimension-one coalescence forms curves in a two-parameter space and is met at isolated values along a generic one-parameter sweep; a codimension-two coalescence occurs only at isolated points of a two-parameter plane. An $m$-fold root at an interior wave number $q_0$ requires $d^jE/dq^j=0$ for $j=1,\dots,m-1$. Because $q_0$ is free, one of these $m-1$ conditions fixes $q_0$; the remaining ones, independent in the generic case, impose $m-2$ conditions on the parameters, so the codimension is $m-2$. An isolated ZGV ($m=2$) is therefore generic, an interior cubic point, where two ZGVs merge, has codimension one, and an interior quartic point, where three ZGVs merge, has codimension two. At a time-reversal-invariant wave number ($q_0=0$ or $\pi/a$), reciprocity makes $E(q)$ even about $q_0$, so $q_0$ is always stationary, all odd derivatives vanish identically, and the multiplicity must be even. With $q_0$ fixed by symmetry, only the even derivatives $j=2,4,\dots,m-2$ must vanish, giving codimension $(m-2)/2$. A ZGV that merges there with its time-reversed partner necessarily joins the preexisting extremum, so three ZGVs coalesce into a quartic root of codimension one. The two lowest-codimension coalescences are thus the interior cubic point and the zone-boundary quartic point. A cubic point cannot occur at a time-reversal-invariant wave number, and the next even case there, $m=6$, has codimension two, like the interior quartic point. The zone-boundary quartic point is therefore the most accessible nontrivial even-multiplicity coalescence. These counts rely on reciprocity; with broken time-reversal symmetry, $E(q)\neq E(-q)$ and the zone boundary loses this status (Sec.~\ref{sec:roots}).

Near an interior coalescence, the local relation reads $A\,\delta E+D\,(\delta q)^3+\dots=0$, the linear and quadratic terms in $\delta q$ having been removed. In a Hermitian lattice $A,D\in\mathbb{R}$, so $\delta q=(-A\,\delta E/D)^{1/3}$ is real for every real $\delta E$, and $E_0$ lies inside the band. At the zone boundary the leading term is quartic, and propagating solutions near $q=\pi/a$ exist on one side of $E_\pi$ only, but other parts of the band can still cover $E_\pi$ (Sec.~\ref{sec:gsp}). In both cases the degeneracy does not by itself remove the continuum. In a non-Hermitian lattice $P$ is complex, two real conditions on $(\delta E,\delta q)$ generically have no real solution, and the degeneracy alone can remove propagating modes; we consider only the Hermitian case.

\section{Root structure, ZGV pairing, and channel counting}
\label{sec:roots}

For a two-leg ladder with reciprocal intrachain couplings of range $r_A$ and $r_B$ and rung coupling $g$, the on-chain dispersions $H_A$ and $H_B$ are polynomials in $c=\cos qa$ because $\cos(nqa)=T_n(c)$. With $E$ and all couplings in units of $\omega_0^2$,
\begin{equation}
	Q(E,c)\equiv[H_A(c)-E][H_B(c)-E]-g^2=0,
	\label{eq:Q}
\end{equation}
with $\deg_cQ=M\equiv r_A+r_B$. The characteristic polynomial $P\propto\lambda^MQ\bigl(E,(\lambda+\lambda^{-1})/2\bigr)$ has degree $2M$ and is palindromic; for the minimal ladder ($r_A=2$, $r_B=1$), $M=3$.

Reciprocity (time-reversal symmetry, class AI) gives $P(E,1/\lambda)=P(E,\lambda)$, which pairs $e^{iqa}$ with $e^{-iqa}$ and has fixed points $\lambda=\pm1$ only. Hermiticity alone gives $P(E,1/\lambda^*)=P(E,\lambda)^*$, which leaves every unimodular root invariant and does not pair $\pm q$. The pairing used here therefore relies on time-reversal symmetry; in a Hermitian lattice with broken time reversal, $E(q)\neq E(-q)$ and the analysis must be redone.

Each root $c$ of $Q$ yields two Bloch factors, $\lambda=c\pm\sqrt{c^2-1}$, the reciprocal solutions of $\lambda+\lambda^{-1}=2c$. A real root in $[-1,1]$ gives a counterpropagating pair $e^{\pm iqa}$. A real root with $|c|>1$ gives a real pair $\lambda,1/\lambda$, alternating in sign for $c<-1$, and a complex-conjugate pair of roots gives a quartet. In either case, each root not in $[-1,1]$ supplies exactly one factor with $|\lambda|<1$, whose amplitude decays as $|\lambda|^n=e^{-na/\xi}$ with $\xi=-a/\ln|\lambda|$. This yields the main-text channel count $d(E)=M-N_{\mathrm{prop}}(E)$, in which each root $c$ is one channel: a propagating channel comprises the pair $e^{\pm iqa}$, and an evanescent channel contributes one factor decaying away from a left boundary. A normalizable state of the semi-infinite lattice is, beyond the boundary layer, a combination of decaying solutions only, so $d(E)\geq1$ is necessary whatever the boundary; it is not sufficient, because the boundary equations must also be solvable within the decaying channels (Sec.~\ref{sec:boundary}).

Because $dc/dq=-a\sin qa$, an interior ZGV obeys $\partial_cE=0$ and is a double root of $Q(E,c)$ in $c$. The two lowest-codimension ways of meeting main-text Eq.~(2) with $k=2$ are as follows. In an interior (saddle-node) coalescence, a neighboring maximum and minimum merge at $q^*\neq0,\pi/a$: $\partial_cE=\partial_c^2E=0$, so $c^*$ is a triple root of $Q$, $\lambda=e^{\pm iq^*a}$ are triple roots of $P$, and $E-E^*\propto(q-q^*)^3$. In zone-boundary self-pairing, $E(q)$ is even about $q=\pi/a$, so $\partial_q^3E|_{\pi/a}=0$ identically, and $\partial_q^2E|_{\pi/a}=-a^2\,\partial_cE|_{c=-1}$; a vanishing inverse effective mass therefore makes $c=-1$ a double root of $Q$. Since $c+1=(\lambda+1)^2/(2\lambda)$, $\lambda=-1$ is then a quadruple root of $P$, and $E-E_\pi\propto(q-\pi/a)^4$. Here an interior ZGV at $q_s$ merges with its mirror image at $2\pi/a-q_s$ and with the extremum at $\pi/a$, so that three ZGVs coalesce. A quartic point in the interior likewise requires three ZGVs to merge at $q^*$, $\partial_cE=\partial_c^2E=\partial_c^3E=0$, so that $c^*$ is a quadruple root of $Q$ and $\lambda=e^{\pm iq^*a}$ are quadruple roots of $P$. Because the evenness of $E(q)$ about $\pi/a$ removes $\partial_q^3E$ there, a quartic point has codimension one at the zone boundary but two in the interior (Secs.~\ref{sec:mult} and \ref{sec:long}).

In both cases the coalesced root is propagating and counts in $N_{\mathrm{prop}}$. In the minimal ladder, an interior coalescence uses all six roots of $P$, so $d(E^*)=0$. At self-pairing, the double root $c=-1$ leaves one further root $c_3$, which for $\eta>1/\sqrt2$ is real and inside $(-1,1)$; at $\eta=1$, $c_3=0$ and $P\propto(\lambda+1)^4(\lambda^2+1)$. Hence $d(E_\pi)=0$, and no normalizable state exists at the coalescence energy.

Near the coalescence the multiple root splits. At the zone boundary, $Q\simeq\alpha(E-E_\pi)+\beta(c+1)^2$ gives $c+1\propto(E-E_\pi)^{1/2}$: for the minimal ladder at $\eta=1$, a complex pair, contributing two decaying channels, for $E<E_\pi$, and a real root below $-1$, contributing one alternating decaying channel, for $E>E_\pi$. Because $c+1\propto(\lambda+1)^2$, the decay rate scales as $1-|\lambda|\propto|E-E_\pi|^{1/4}$. At an interior saddle-node, $Q\propto(c-c^*)^3$ splits on both sides of $E^*$ into one real and two complex roots, with $1-|\lambda|\propto|E-E^*|^{1/3}$. The real root stays in $[-1,1]$, so each of the two decaying factors, near $e^{\pm iq^*a}$, has a propagating factor within $\sim a/\xi$ of it; in the minimal ladder these are all the channels. At the zone boundary, by contrast, both decaying factors lie near $\lambda=-1$, and the propagating channel $c_3$ lies far from it. Section~\ref{sec:locking} shows that this difference decides whether a resonance forms.

Below criticality, $g<\gsp$, the quadruple root unfolds (Sec.~\ref{sec:gsp}). The zone boundary becomes a weak local maximum, $c=-1$ is a simple root of $Q(E_\pi,c)$, and the interior minimum at $q_s$ and its mirror image lie at $E_s<E_\pi$, where $c_s=\cos q_sa$ is the double root of an ordinary interior ZGV. In the narrow window $E_s<E<E_\pi$ the two roots near $c=-1$ are real and in $[-1,1]$, so $d=0$. Below $E_s$ the double root $c_s$ splits into a complex pair close to $c=-1$, giving $d=2$, the same channel content as just below $E_\pi$ at $g=\gsp$. Its decay rate crosses over from $1-|\lambda|\propto(E_s-E)^{1/2}$, the law of an isolated ZGV, for $E_s-E\ll E_\pi-E_s$, to the quartic law for $E_s-E\gtrsim E_\pi-E_s$. Because $E_\pi-E_s$ vanishes at criticality, the split roots remain close to the unit circle near $E_\pi$ slightly below criticality, and the weakly evanescent channels persist.

\section{Two-leg ladder model}
\label{sec:model}

With $\omega_0^2=k_1/m$, $\eta=k_2/k_1$, $g=k_c/k_1$, and $k_w=2k_2$, the displacements $u_{nA}$ and $u_{nB}$ of the $n$th masses on chains A and B obey
\begin{align}
	\omega^2u_{nA}&=(2+2\eta+g)\omega_0^2u_{nA}-\omega_0^2(u_{n+1,A}+u_{n-1,A})\notag\\
	&\quad-\eta\omega_0^2(u_{n+2,A}+u_{n-2,A})-g\omega_0^2u_{nB},
	\label{eq:eomA}\\
	\omega^2u_{nB}&=(2+2\eta+g)\omega_0^2u_{nB}-\omega_0^2(u_{n+1,B}+u_{n-1,B})\notag\\
	&\quad-g\omega_0^2u_{nA}.
	\label{eq:eomB}
\end{align}
In the bosonic ladder of Ref.~\cite{downing2024}, on-site energies are set independently of the hoppings. In the phononic ladder every spring contributes both a coupling and a diagonal self-stiffness, and $k_w=2k_2$ is the unique grounding that equalizes the bulk self-stiffnesses, $2k_1+2k_2+k_c$ on both chains.

Because $E=\omega^2$ and every coalescence considered here occurs at nonzero frequency, $E-E_0\simeq2\omega_0(\omega-\omega_0)$ to leading order, so energy and frequency detunings obey the same scaling exponents. With $E=\omega^2/\omega_0^2$, Eqs.~\eqref{eq:eomA} and \eqref{eq:eomB} are the eigenproblem of a two-leg tight-binding Hamiltonian with on-site energy $2+2\eta+g$, hoppings $-1$ along both chains, $-\eta$ between next-nearest neighbors on chain A, and $-g$ on the rungs. Nothing in Secs.~\ref{sec:mult}, \ref{sec:roots}, \ref{sec:gsp}, \ref{sec:locking}, and \ref{sec:long} uses the phononic origin of these equations, so the multiplicity conditions, the channel count, and $\gsp(\eta)$ hold for any Hermitian, time-reversal-symmetric lattice with the same hoppings, with $\omega^2$ replaced by $E$. Examples are coupled-mode equations of zigzag waveguide arrays, with $E$ the propagation constant~\cite{szameit2009zigzag}; polariton micropillar lattices~\cite{amo2016polariton}; and electrical circuits, whose Laplacian has the structure of a dynamical matrix~\cite{lee2018topolectrical}. An overall sign or shift of the Hamiltonian maps the upper band onto the lower one and leaves the analysis unchanged.

In Bloch form, $\omega_\pm^2=\omega_0^2(\Sigma\pm R)$ with
\begin{equation}
	\Sigma=2+2\eta+g-2\cos qa-\eta\cos2qa,\quad
	R=\sqrt{\Delta^2+g^2},
	\label{eq:bands}
\end{equation}
and $\Delta=-\eta\cos2qa$, half the difference of the bare chain dispersions. The bare bands cross at $q_ca=3\pi/4$, where chain B has positive group velocity and chain A has $v_g^A\propto\sqrt2-4\eta$. Counterpropagating bare bands therefore require $\eta>1/(2\sqrt2)$. The sign of $\Sigma'(q_c)\propto\sqrt2-4\eta$ then determines the side of the crossing on which hybridization creates the interior ZGV of the upper band: below $q_c$ for $1/(2\sqrt2)<\eta<1/\sqrt2$, and above $q_c$, on the zone-boundary side, for $\eta>1/\sqrt2$, from where it migrates toward $qa=\pi$ as $g$ increases.

\section{Self-pairing coupling}
\label{sec:gsp}

With $F(q)=\omega_+^2(q)$, $F'=\omega_0^2(\Sigma'+\Delta\Delta'/R)$, where $\Sigma'=2a(\sin qa+\eta\sin2qa)$ and $\Delta'=2a\eta\sin2qa$. Setting $F'=F''=0$, substituting $c=\cos qa$, and eliminating $g$ gives
\begin{equation}
	6\eta^2c^4+6\eta c^3+(1-5\eta^2)c^2-6\eta c-(1+\eta^2)=0,
	\label{eq:quartic}
\end{equation}
which factors as
\begin{equation}
	(c^2-1)\left(6\eta^2c^2+6\eta c+1+\eta^2\right)=0.
	\label{eq:factor}
\end{equation}
The quadratic factor has real roots only for $\eta\leq1/\sqrt2$ and describes interior saddle-nodes, at which $d=0$ (Sec.~\ref{sec:roots}); they are physical, with $c^*\in(-1,1)$ and real $g^*>0$, for $1/(2\sqrt2)<\eta\leq1/\sqrt2$, with $g^*\to0$ at the lower end (e.g., $g^*=0.157$, $0.404$, and $1.093$ at $\eta=0.5$, $0.6$, and $0.7$). These cubic points produce no resonance (Sec.~\ref{sec:locking}). The factor $c=+1$ would describe self-pairing at the zone center, which cannot occur (Sec.~\ref{sec:long} with $p=2$). For $c=-1$, $\Delta'=0$ and
\begin{equation}
	F''(\pi/a)=a^2\omega_0^2\left[(4\eta-2)-\frac{4\eta^2}{\sqrt{\eta^2+g^2}}\right],
	\label{eq:Fpp}
\end{equation}
whose zero gives main-text Eq.~(3). At the zone boundary, where $d\omega_+/dq=0$, the inverse effective mass defined in the main text is $1/\meff=d^2\omega_+/dq^2=F''/(2\omega_+)$; it changes sign from negative ($g<\gsp$, local maximum) through zero ($g=\gsp$) to positive ($g>\gsp$, local minimum).

The expression for $\gsp$ is real and positive for all $\eta>1/2$, but its character changes at two thresholds set by the quadratic factor of Eq.~\eqref{eq:factor}. First, that factor passes through $c=-1$ when $7\eta^2-6\eta+1=0$, at $\eta_c=(3+\sqrt2)/7\approx0.63$, where $F''$ and $F''''$ vanish together at $\pi/a$. For $1/2<\eta<\eta_c$, $F''''(\pi/a)<0$ at $g=\gsp$, so the self-pairing point is a quartic maximum at the top of the upper band and is not embedded in the continuum; for $\eta>\eta_c$ it lies inside the band. Second, for $\eta_c<\eta<1/\sqrt2$, interior saddle-nodes still coexist with self-pairing, and the interior minimum reaches $qa=\pi$ only within a few percent of $\gsp$. For $\eta>1/\sqrt2$ the quadratic factor has no real roots, and self-pairing is the only degenerate stationary point of the upper band. We therefore state main-text Eq.~(3) for $\eta>1/\sqrt2$.

For $\eta>1/\sqrt2$ and $g<\gsp$, the upper band has an interior maximum near $qa\approx1.9$, an interior minimum at $q_s$, and a local maximum at $\pi/a$. As $g\to\gsp$, the minimum at $q_s$ and its mirror image converge on $\pi/a$ and merge with the zone-boundary maximum, where $F'=F''=F'''=0$ and $F''''\neq0$; for $g>\gsp$, $\pi/a$ is a single minimum. The interior maximum persists throughout, so $E_\pi$ remains embedded in the band [main-text Fig.~1(b)]. At $\eta=1$ and $g=\gsp$, the maximum lies at $qa\approx1.94$, about 2\% above $\omega_\pi=2.955\,\omega_0$, and near the zone boundary
\begin{equation}
	E-E_\pi\simeq0.50\,\omega_0^2\,[(q-\pi/a)a]^4 .
	\label{eq:Equartic}
\end{equation}

The pocket of low $\PR_{\min}$ in main-text Fig.~3(a) lies above $\eta=1/\sqrt2$, consistent with these thresholds and with the branch selection of Sec.~\ref{sec:model}. It becomes pronounced once the self-pairing locus enters the regime in which the upper and lower bands overlap; at $g=\gsp(\eta)$, the gap between the upper-band minimum and the lower-band maximum closes at $\eta\simeq0.8$. Throughout the pocket, the minimum-$\PR$ mode lies inside the upper band close to $\omega_+(\pi/a)$, and the upper edge of the pocket approximately follows $\gsp(\eta)$.

\section{Boundary problem}
\label{sec:boundary}

Cutting the intercell bonds at $n=1$ removes the couplings to the fictitious sites $u_{0A}$, $u_{-1,A}$, and $u_{0B}$. In the tight-binding lattice the on-site terms are unchanged; in the phononic lattice, cutting a bond also removes its diagonal contribution, which grounding springs $k_1+k_2$, $k_2$, and $k_1$ on sites $1A$, $2A$, and $1B$ (and their mirror images at $n=N$) restore. With this bulk-matched termination, the three boundary equations reduce to the vanishing of the missing coupling terms. For $\psi_n=\sum_ja_j\lambda_j^n\,(e_A^{(j)},e_B^{(j)})^{T}$ these are
\begin{align}
	\textstyle\sum_ja_je_A^{(j)}\bigl(1+\eta\lambda_j^{-1}\bigr)&=0,\label{eq:bc1}\\
	\textstyle\sum_ja_je_A^{(j)}&=0,\label{eq:bc2}\\
	\textstyle\sum_ja_je_B^{(j)}&=0,\label{eq:bc3}
\end{align}
from sites $1A$, $2A$, and $1B$, respectively. For free ends, the diagonal deficit adds terms proportional to $u_{1A}$, $u_{2A}$, and $u_{1B}$, an edge potential. More generally, cutting a ladder with coupling ranges $r_A$ and $r_B$ removes $r_A+r_B=M$ fictitious sites beyond the cut, so the boundary problem comprises $M$ equations, one per missing site; for the longer-range ladder of Sec.~\ref{sec:long}, these are six on chain A and two on chain B.

A boundary state at energy $E$ may use only the $d(E)$ decaying factors, so Eqs.~\eqref{eq:bc1}--\eqref{eq:bc3} form a $3\times d$ linear system; in general the system is $M\times d$, and since $d=M-N_{\mathrm{prop}}\le M$, a nontrivial solution at a single free energy generically requires $d=M$, i.e., no propagating channel. At $E_\pi$ with $g=\gsp$, $d=0$ and no solution exists. Just below $E_\pi$, where the resonance forms, $d=2$, and a nontrivial solution requires the $3\times2$ matrix to be rank deficient, which the bulk degeneracy does not enforce. It does, however, make the matrix nearly rank deficient: both decaying factors converge on $\lambda=-1$ as $E\to E_\pi$, so their columns become nearly parallel. With normalized columns, the smaller singular value is 0.09 at $E-E_\pi=-10^{-3}\omega_0^2$ and 0.05 at $-10^{-4}\omega_0^2$, against 0.65--0.8 at the minimal-ladder saddle-nodes ($\eta=0.5$--0.7), whose two decaying factors lie near the distinct roots $e^{\pm iq^*a}$. The near-self-pairing mode is therefore built mainly from the two weakly evanescent channels with a residual admixture of the propagating channel, a resonance rather than a bound state; Sec.~\ref{sec:locking} makes this quantitative. The factor $1+\eta\lambda^{-1}$ in Eq.~\eqref{eq:bc1} vanishes at $\lambda=-\eta$, which lies on the unit circle at $\eta=1$ and so cannot underlie the boundary concentration found there.

Because the resonance relies on the boundary equations above, it is sensitive to the termination. At $g=\gsp$ and $N=400$, adding 5\% of the free-end edge potential to the terminal sites raises $\PR_{\min}$ from 0.10 to 0.15 and lowers $W_b$ from 0.59 to 0.43; a uniform on-site detuning of a few percent of $k_1$ has a comparable effect. For free ends, the minimum-$\PR$ mode is a lower-band standing wave with the uniform boundary weight. The enhancement near $E_\pi$ is not destroyed but weakened: in the semi-infinite lattice its evanescent weight $W$ [Eq.~\eqref{eq:W}] still grows as $\xi^3$ for free ends, with about half the prefactor of the bulk-matched termination (Sec.~\ref{sec:locking}), which at accessible $N$ is not enough to make it the minimum-$\PR$ mode. The termination thus sets the strength of the resonance, whereas the bulk sets its scaling.

\section{Locking at cubic coalescences}
\label{sec:locking}

This section shows why a cubic coalescence produces no boundary resonance, whereas a quartic one does, and why the conclusion is independent of the termination.

\textit{Root clusters.}---Near an $m$-fold root $\lambda_0=e^{iq_0a}$ on the unit circle, the dispersion relation reduces to $A\,\delta E+D\,(\delta q)^m=0$, with $A,D\in\mathbb{R}$ (Sec.~\ref{sec:mult}), so
\begin{equation}
	\delta q=\left(-A\,\delta E/D\right)^{1/m}e^{2\pi ik/m},\qquad k=0,\ldots,m-1 .
	\label{eq:cluster}
\end{equation}
The $m$ roots form a cluster of radius $\sim a/\xi$ around $q_0$. Roots with $\operatorname{Im}\delta q>0$ decay away from a left termination. For odd $m$, one root of the cluster is real on both sides of $E_0$, and $(m-1)/2$ roots decay. For even $m$, two roots are real and $(m-2)/2$ decay on one side of $E_0$, while on the other side $m/2$ roots decay and none is real. Applied to the coalescences of Sec.~\ref{sec:roots}:
\begin{itemize}
	\item at an interior saddle-node ($m=3$ at $e^{\pm iq^*a}$), each of the two clusters contains one decaying factor and one propagating factor;
	\item at a quartic point ($m=4$, at $\lambda=-1$ for self-pairing or at $e^{\pm iq^*a}$ for an interior three-ZGV merger), each cluster contains, on its evanescent side, two decaying factors and no propagating one;
	\item at an isolated ZGV ($m=2$), each cluster contains, on its evanescent side, one decaying factor and no propagating one.
\end{itemize}

\textit{Boundary response.}---Consider the semi-infinite lattice with a wave incident in one propagating channel. The $M$ boundary equations (Sec.~\ref{sec:boundary} for $M=3$) involve the column vectors $\mathbf{b}(\lambda)$ obtained by inserting $\lambda^n\mathbf{e}(\lambda)$, and the $M$ unknowns are the $d$ decaying and $M-d$ outgoing amplitudes. Within a cluster, the columns differ by $O(a/\xi)$. We therefore expand $\mathbf{b}(\lambda_j)=\mathbf{b}(\lambda_0)+(\lambda_j-\lambda_0)\mathbf{b}'(\lambda_0)+\dots$ and collect the cluster amplitudes into divided-difference coefficients, and we assume that the resulting matrix, built from $\mathbf{b}(\lambda_0)$, $\mathbf{b}'(\lambda_0)$, and the columns of the other channels, has full rank, which holds generically. Here a generic local termination is any termination acting on finitely many cells that imposes the same $M$ boundary conditions and for which this matrix has full rank; terminations tuned to violate this condition, or that add boundary degrees of freedom, are excluded.

At a cubic coalescence, let cluster 1 contain the incident factor $\lambda_{\mathrm{in}}$ (amplitude 1) and the decaying factor $\lambda_1$ (amplitude $a_1$), and cluster 2 the outgoing factor $\lambda_{\mathrm{out}}$ (amplitude $r$) and the decaying factor $\lambda_2$ (amplitude $a_2$). The boundary equations become
\begin{equation}
	x_1\mathbf{b}(\lambda_{\mathrm{in}})+x_2\mathbf{b}'(\lambda_{\mathrm{in}})+x_3\mathbf{b}(\lambda_{\mathrm{out}})+x_4\mathbf{b}'(\lambda_{\mathrm{out}})+\dots=0,
\end{equation}
with $x_1=1+a_1$, $x_2=a_1(\lambda_1-\lambda_{\mathrm{in}})$, $x_3=r+a_2$, and $x_4=a_2(\lambda_2-\lambda_{\mathrm{out}})$. In the minimal ladder these are three equations in four coefficients, whose solutions form a one-parameter family $(x_1,\dots,x_4)=t\,\mathbf{v}$. The relation $x_1=1+x_2/(\lambda_1-\lambda_{\mathrm{in}})$ then fixes $t=O(a/\xi)$, so
\begin{equation}
	a_1=-1+O(a/\xi),\qquad a_2=-r+O(a/\xi).
	\label{eq:lock}
\end{equation}
Each decaying amplitude locks to $-1$ times the propagating amplitude in its own cluster; with further channels (larger $M$) the count of equations and coefficients rises equally, and the conclusion is unchanged. Near the termination,
\begin{equation}
	\psi_n\simeq\lambda_p^n-\lambda_d^n=\lambda_p^n\left[1-e^{i(q_d-q_p)n a}\right]\propto n a/\xi,\qquad n\ll\xi/a,
	\label{eq:deplete}
\end{equation}
so the evanescent part fills in the near field of an ordinary reflection, and the boundary is depleted.

\textit{Evanescent weight.}---We normalize each polarization vector, $|\mathbf{e}^{(j)}|=1$, so that the incident wave of unit amplitude carries unit intensity per cell. Writing the evanescent part of the scattering solution as $\psi_n^{\mathrm{ev}}=\sum_{j\in\mathrm{dec}}a_j\lambda_j^n\mathbf{e}^{(j)}$, a sum over the $d$ decaying factors, we define the evanescent weight of the semi-infinite lattice as
\begin{equation}
	W=\sum_{n\geq1}\bigl|\psi_n^{\mathrm{ev}}\bigr|^2,
	\label{eq:W}
\end{equation}
the weight carried by the evanescent channels in units of the bulk intensity per cell. It is distinct from the finite-lattice fraction $W_b$ of Eq.~\eqref{eq:wb}, which is normalized and bounded by unity. At a cubic point, Eq.~\eqref{eq:lock} gives $|a_1|\to1$, so the norm of the evanescent component tends to $W\simeq\sum_{n\geq1}e^{-2na/\xi}\simeq\xi/(2a)$, with a universal prefactor that does not depend on the termination. This component, however, interferes destructively with the propagating component near the boundary, so a large $W$ does not imply a large boundary intensity: the boundary is depleted.

At a quartic point, the two decaying factors of a cluster enter as $x_1=a_1+a_2$ and $x_2=a_1(\lambda_1-\lambda_0)+a_2(\lambda_2-\lambda_0)$, both $O(1)$ for a generic termination. The evanescent field is then $x_1\lambda_0^n+x_2\,(\lambda_1^n-\lambda_2^n)/(\lambda_1-\lambda_2)$, and the second term behaves as $n\lambda_0^{n-1}e^{-na/\xi}$, so
\begin{equation}
	W\propto\sum_n n^2e^{-2na/\xi}\propto(\xi/a)^3,\qquad W/\xi\propto|\delta E|^{-1/2}.
	\label{eq:xi3}
\end{equation}
At an isolated ZGV the single decaying factor has an $O(1)$, termination-dependent amplitude, and $W\propto\xi$. The same counting suggests $W\propto\xi^{m-1}$ for any even $m$, since $m/2$ decaying factors combine into profiles up to $n^{m/2-1}\lambda_0^ne^{-na/\xi}$. For a finite ladder whose level nearest the coalescence lies at $|\delta E|\propto N^{-1}$, the boundary enhancement over the uniform value is then expected to grow asymptotically as $N^{3/4}$ near a quartic point, against $N^{1/2}$ near an isolated ZGV.

\textit{Numerical check.}---We solved the semi-infinite scattering problem exactly for four terminations: bulk matched, bulk matched plus 5\% of the free-end edge potential, a random edge potential, and free ends. At the minimal-ladder saddle-nodes ($\eta=0.5$, 0.6, and 0.7 at their $g^*$), $W/\xi\to0.50\pm0.02$ on both sides of $E^*$ for every termination, and the mean intensity within $\xi/4$ of the termination is below 11\% of its bulk value. At self-pairing ($\eta=1$, $g=\gsp$), the local exponent of $W$ versus $\xi$ approaches 3 for every termination, with $W/\xi^3\to0.78$ for the bulk-matched termination and $0.37$ for free ends: the termination sets the prefactor, not the scaling.

\begin{figure*}[t]
	\includegraphics[width=\textwidth]{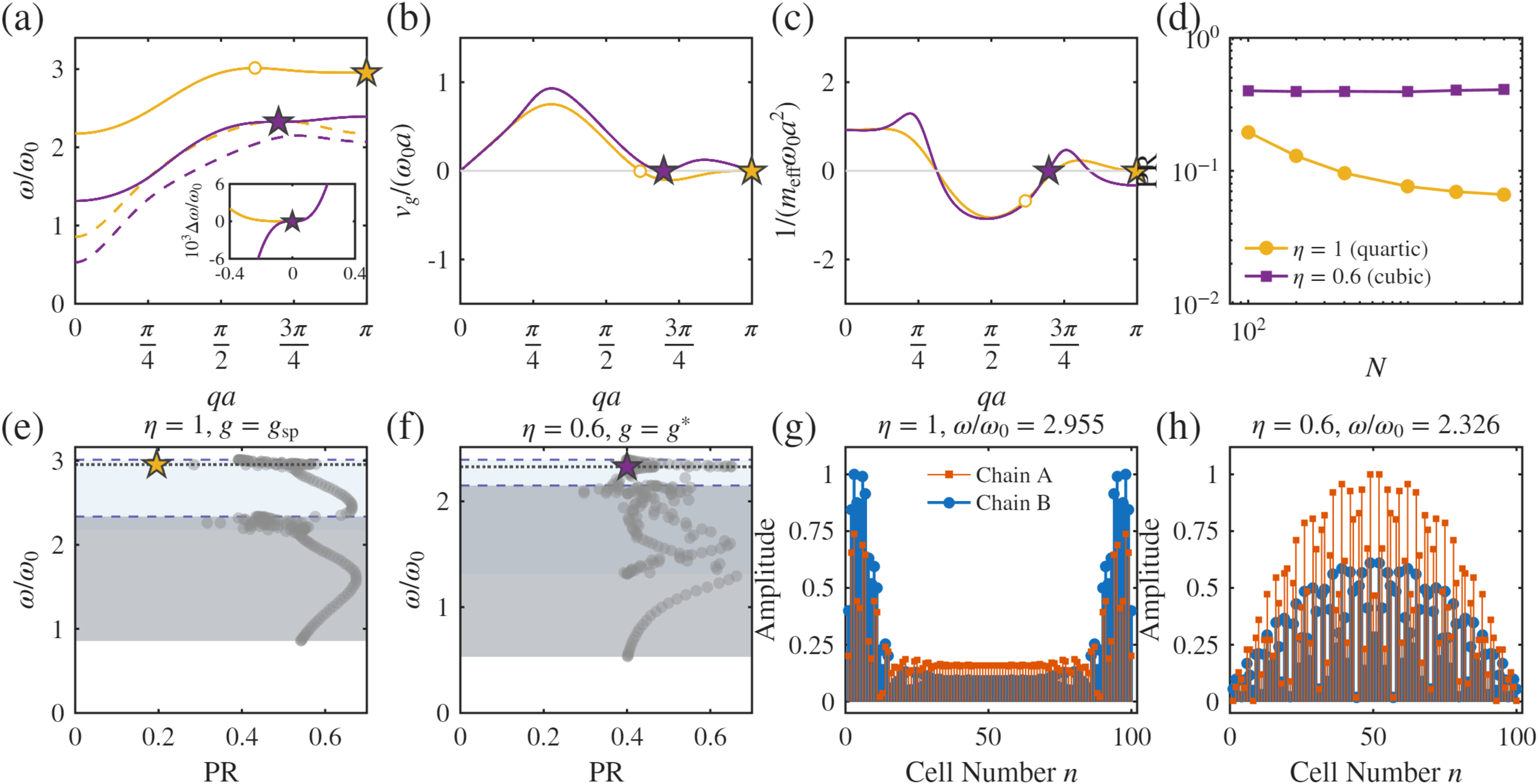}
	\caption{\label{fig:Smin}
		Quartic versus cubic coalescence in the minimal ladder: zone-boundary self-pairing at $\eta=1$, $g=\gsp$ (yellow) and interior saddle-node at $\eta=0.6$, $g=g^*=0.404$ (purple). (a) Bands (upper solid, lower dashed); inset: $10^3(\omega-\omega_c)/\omega_0$ versus $(q-q_c)a$ around the coalescence. (b) Group velocity and (c) inverse effective mass of the upper band; open circles mark upper-band ZGVs and stars the coalescences. (d) $\PR$ of the selected mode versus $N$. (e),(f) $\PR$ versus $\omega/\omega_0$ for $N=100$ and (g),(h) normalized amplitudes of the selected modes: the minimum-$\PR$ mode within $0.05\,\omega_0$ below $\omega_\pi$ (yellow) and within $0.05\,\omega_0$ of $\omega^*$ (purple), narrowed to $0.01\,\omega_0$ for $N>400$.}
\end{figure*}

Figure~\ref{fig:Smin} compares the two cases in finite ladders. At self-pairing, $\PR$ of the resonance falls from 0.195 at $N=100$ to 0.066 at $N=4000$, while $W_b$ stays 4--40 times its uniform value. Near the saddle-node at $\eta=0.6$, the minimum-$\PR$ mode within $0.05\,\omega_0$ of $\omega^*$ has $\PR=0.39$--0.41 at every $N$ and $W_b\leq0.03$, below the uniform value, with the depleted edges of Eq.~\eqref{eq:deplete} [Fig.~\ref{fig:Smin}(h)]. Levels accumulate at $E^*$ because the density of states diverges there, and the selected mode converges to $E^*$ with vanishing edge weight.

\section{Finite-size resonance}
\label{sec:finite}

\textit{Boundary measures.}---With $u_{nA}$ and $u_{nB}$ the amplitudes of the $n$th masses, the boundary weight is the fraction of modal weight in the $n_b$ cells nearest the two ends,
\begin{equation}
	W_b=\frac{\sum_{n\in\mathcal{B}}\left(|u_{nA}|^2+|u_{nB}|^2\right)}{\sum_{n=1}^{N}\left(|u_{nA}|^2+|u_{nB}|^2\right)},
	\label{eq:wb}
\end{equation}
with $\mathcal{B}=\{1,\ldots,n_b\}\cup\{N-n_b+1,\ldots,N\}$, so a uniform mode has $W_b=2n_b/N$; we use $n_b=10$ for the minimal ladder and $n_b=20$ for the longer-range ladder. The participation ratio is
\begin{equation}
	\PR=\frac{\left[\sum_n\left(|u_{nA}|^2+|u_{nB}|^2\right)\right]^2}{2N\sum_n\left(|u_{nA}|^4+|u_{nB}|^4\right)},
	\label{eq:pr}
\end{equation}
which equals unity for a uniform mode; $\PR_{\min}$ is its minimum over all eigenmodes.

A state bound to the boundary with a fixed decay length has a saturating number of participating sites, so $\PR_{\min}\propto N^{-1}$, whereas an extended state has $\PR_{\min}\to$ const. Main-text Fig.~3(b) shows $\PR_{\min}$ of the bulk-matched minimal ladder at $\eta=1$, with $\PR$ defined in Eq.~\eqref{eq:pr}; for $N\geq2000$ we compute only the eigenpairs near the minimum-$\PR$ mode by shift-invert Lanczos. Away from self-pairing ($g/\gsp\le0.7$ or $\geq1.1$), $\PR_{\min}$ stays between 0.15 and 0.45 without decreasing. For $0.95\lesssim g/\gsp\lesssim1.05$ it decreases with $N$ without following a power law: at $g=\gsp$ it falls from 0.20 at $N=100$ to 0.07 at $N=4000$, while the local exponent $-d\ln\PR_{\min}/d\ln N$ falls from about 0.6 to below 0.1. Over the same range $W_b$ falls from 0.83 to 0.20, 40 times the uniform value $20/N$. At the largest $N$, the level spacing near $\omega_\pi$ becomes small enough that the boundary weight is shared between two nearly degenerate eigenmodes rather than concentrated in one; at $g=0.95\,\gsp$ this raises $\PR_{\min}$ to $0.12$ at $N=5000$, the upturn in main-text Fig.~3(b). For comparison, a stiff grounding spring $k_{\mathrm{def}}=3k_1$ added to the terminal chain-A sites at $g=0.95\,\gsp$ creates a pair of degenerate end states at $\omega=3.145\,\omega_0$, above the upper band; resolved by reflection parity, each has $\PR\propto N^{-1.00}$.

The channel content follows from the roots of $Q$ at the eigenfrequency $\omega_{\mathrm{res}}$ of the minimum-$\PR$ mode. At $\eta=1$, $g=\gsp$, and $N=100$, $\omega_{\mathrm{res}}-\omega_\pi=-3.0\times10^{-4}\,\omega_0$, where
\begin{equation}
	c_{1,2}\simeq-1\pm0.030\,i,\qquad c_3\simeq0.001 .
	\label{eq:roots}
\end{equation}
Only $c_3$ is propagating, at $qa\simeq\pi/2$, so $d=2$; the complex pair gives $\lambda_{1,2}\simeq-0.83\pm0.14\,i$ with $\xi/a\simeq5.8$, both descending from the quadruple root $\lambda=-1$. As $N$ increases, $\omega_{\mathrm{res}}$ approaches $\omega_\pi$ from below with $d=2$ throughout; only exactly at $E_\pi$ does the pair collapse onto $c=-1$. From $N=100$ to 2000 the detuning shrinks 3.2-fold, to $-9.4\times10^{-5}\,\omega_0$, and $\xi/a$ grows only from 5.8 to 7.7, consistent with $\xi\propto|E-E_\pi|^{-1/4}$, so $W_b$ decreases while the channel content is unchanged.

Below criticality, the minimum-$\PR$ mode lies just below the interior minimum $E_s$, outside the window $E_s<E<E_\pi$ in which $d=0$, and has the same channel content: a complex pair split from $c_s$ close to $c=-1$ and a propagating pair near $qa\simeq\pi/2$ (Sec.~\ref{sec:roots}). For $0.9\lesssim g/\gsp<1$, its $\PR_{\min}$ and $W_b$ remain close to those at $\gsp$; at $g/\gsp\simeq0.8$--0.85 the mode is still boundary weighted at $N=100$ but weakens with increasing $N$, consistent with the upper edge of the low-$\PR$ pocket.

The resonance is not a standing wave of the flat branch. Such standing waves, quantized at $|q-\pi/a|\approx\pi/(Na)$, are detuned from $E_\pi$ by $\propto N^{-4}$ and, because Eq.~\eqref{eq:Equartic} is a quartic minimum, lie above $E_\pi$; between $N=100$ and 2000 their detuning would shrink by about $10^5$. The resonance instead lies below $E_\pi$, where the flat branch is evanescent, and its detuning shrinks only 3.2-fold. Its frequency is set by matching the weakly evanescent channels to the termination, not by quantizing the flat branch.

\section{Longer-range couplings and interior coalescence}
\label{sec:long}

For a single coupling $k_p$ of range $p\geq2$ on chain A, with grounding $2k_p$ on chain B, the bands are $\omega_\pm^2=\Sigma\pm R$ with $\Sigma=2k_1+2k_p+k_c-2k_1\cos qa-k_p\cos pqa$ and $R=\sqrt{k_c^2+k_p^2\cos^2pqa}$ (units $m=1$), and $M=p+1$. The zone boundary is always stationary, with
\begin{equation}
	\partial_q^2\Sigma\big|_{\pi/a}=a^2[-2k_1+p^2k_p(-1)^p],\qquad
	\partial_q^2R\big|_{\pi/a}=-\frac{a^2p^2k_p^2}{\sqrt{k_c^2+k_p^2}},
	\label{eq:Sp}
\end{equation}
so self-pairing at $\pi/a$ requires $p^2k_p^2/\sqrt{k_c^2+k_p^2}=p^2k_p(-1)^p-2k_1$. For odd $p$ the right-hand side is negative and there is no solution. For even $p$ with $p^2k_p>2k_1$,
\begin{equation}
	\gsp^{(p)}(r)=r\sqrt{\frac{p^4r^2}{(p^2r-2)^2}-1},\qquad r=k_p/k_1,
	\label{eq:gspp}
\end{equation}
which reduces to main-text Eq.~(3) for $p=2$ and to $2\sqrt{p^2-1}/(p^2-2)$ for $k_p=k_1$. The zone center never self-pairs: the analogous condition, $2k_1+p^2k_p=p^2k_p^2/\sqrt{k_c^2+k_p^2}$, cannot hold because its right-hand side is at most $p^2k_p$. For $M\leq3$, an interior saddle-node uses all $2M$ Bloch roots, so at nearby energies its split roots, a locked evanescent--propagating pair in each cluster, are the only channels (Secs.~\ref{sec:roots} and \ref{sec:locking}); for longer ranges, $F'=F''=0$ determines $(q^*,k_c^*)$ numerically, and additional channels can be present.

For the asymmetric ladder of the main text, $k_{1A}\equiv k_1$, $k_{1B}=2k_1$, $k_{6A}=3k_1$, and $k_w=2k_{1A}+2k_{6A}-2k_{1B}-2k_{2B}$, so that both chains have the bulk self-stiffness $S=2k_{1A}+2k_{6A}+k_c$:
\begin{align}
	H_A&=S-2k_{1A}\cos qa-2k_{6A}\cos6qa,\notag\\
	H_B&=S-2k_{1B}\cos qa-2k_{2B}\cos2qa,
	\label{eq:asym}
\end{align}
and $M=8$. The bulk-matched termination cuts the bonds and keeps $S$ on every site, as in Sec.~\ref{sec:boundary}.

\textit{Cubic coalescence.}---At $k_{2B}=1.1k_1$ ($k_w=1.8k_1$), the upper band has, below $k_c^*$, a closely spaced minimum--maximum pair near $qa\approx1.92$--2.00 and a further minimum near $qa\approx2.26$. The pair coalesces at $k_c=1.059k_1$, $qa=1.965$, and $\omega=3.509\,\omega_0$, where all eight $c$ roots of $Q$, including the triple root $c=-0.384$, are real and in $[-1,1]$, so $d(E)=0$. Above $k_c=1.059k_1$ only the outer minimum remains on this segment. At $N=500$ the minimum-$\PR$ mode lies at $3.494\,\omega_0$, just below the adjacent minimum at $3.4985\,\omega_0$, with $\PR=0.087$ and $W_b=0.47$. Its channel content, $d=4$, consists of a decaying pair split from the coalescence ($\operatorname{Re}qa\simeq1.85$, $\xi/a\simeq9$), which keeps its propagating partners and is locked (Sec.~\ref{sec:locking}), and a pair from the split double root of the adjacent minimum ($\operatorname{Re}qa\simeq2.27$, $\xi/a\simeq16$), which has none. The latter carries the boundary weight. A mode of the same kind, just below that minimum, appears for $0.5\leq k_c/k_1\lesssim1.7$, including $k_c<1.059k_1$, where the pair has not yet coalesced, and $k_c\geq1.2k_1$, where only the minimum remains, with no systematic change at $k_c=1.059k_1$. Its $\PR$ levels off with $N$, at 0.040 for $N=4000$ at $k_c=1.059k_1$, as expected for an isolated ZGV.

\begin{figure*}[t]
	\includegraphics[width=\textwidth]{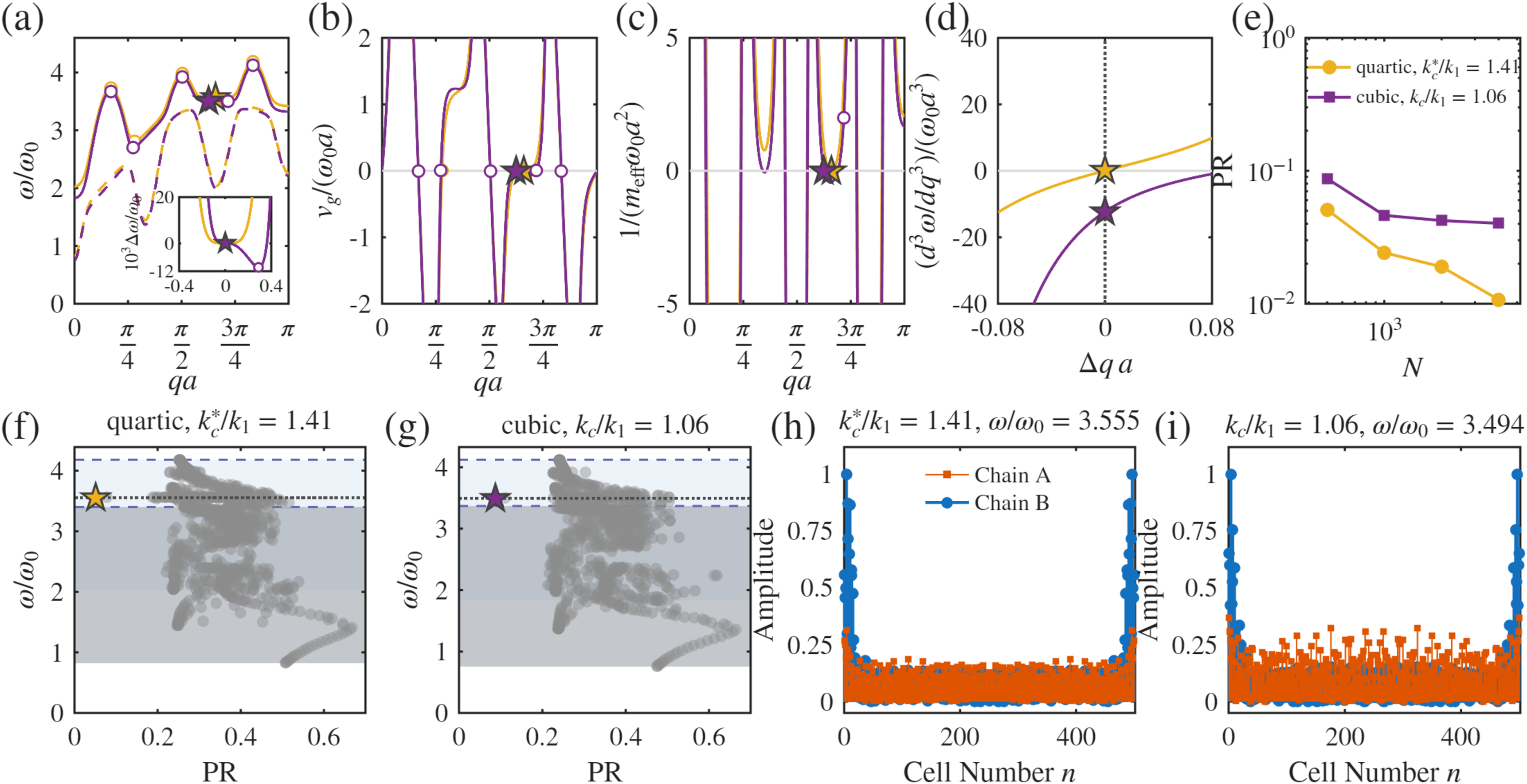}
	\caption{\label{fig:SM8}
		Quartic versus cubic coalescence in the longer-range ladder: interior three-ZGV merger at $k_c^*/k_1=1.41$, $k_{2B}/k_1=1.014$ (yellow) and two-ZGV coalescence at $k_c/k_1=1.06$, $k_{2B}/k_1=1.1$ (purple). (a) Bands (upper solid, lower dashed); inset: $10^3(\omega-\omega_c)/\omega_0$ versus $(q-q_c)a$ around the coalescence. (b) Group velocity and (c) inverse effective mass of the upper band; open circles mark upper-band ZGVs and stars the coalescences. (d) Third derivative $d^3\omega/dq^3$ of the upper band versus $(q-q_c)a$: it vanishes at the quartic point and not at the cubic one. (e) $\PR$ of the selected mode versus $N$. (f),(g) $\PR$ versus $\omega/\omega_0$ for $N=500$ and (h),(i) normalized amplitudes of the selected modes. The selected mode is the minimum-$\PR$ mode within $0.05\,\omega_0$ ($0.01\,\omega_0$ for $N>500$) below the quartic point (yellow) or below the adjacent ordinary minimum (purple).}
\end{figure*}

\textit{Interior quartic point.}---Tuning $k_{2B}$ as a second parameter moves the minimum onto the coalescence. Requiring $Q=\partial_cQ=\partial_c^2Q=\partial_c^3Q=0$ at $(E^*,c^*)$ gives $k_c^*=1.408k_1$, $k_{2B}=1.014k_1$ ($k_w=1.972k_1$), $c^*=-0.479$, $q^*a=2.070$, and $\omega^*=3.558\,\omega_0$. The upper band then has an interior quartic minimum embedded in the band, at which $d^3\omega/dq^3$ vanishes together with the group velocity and inverse effective mass, whereas at the cubic point it remains finite, $d^3\omega/dq^3=-12.4\,\omega_0a^3$ [Fig.~\ref{fig:SM8}(d)], and just below $\omega^*$ all four decaying factors ($\operatorname{Re}qa\simeq1.95$ and $2.20$) descend from the quadruple roots $e^{\pm iq^*a}$, whereas the four propagating channels, at $qa\simeq0.42$, 0.64, 1.39, and 2.89, lie outside the clusters. Figure~\ref{fig:SM8} compares it with the cubic point. The quartic resonance has $\PR=0.051$, 0.024, 0.019, and 0.011 at $N=500$, 1000, 2000, and 4000, with $W_b$ 8, 17, 25, and 33 times its uniform value, whereas the cubic case levels off at $\PR\simeq0.04$, with $W_b$ 6--21 times uniform. At fixed $k_{2B}=1.014k_1$, the resonance persists for $1\lesssim k_c/k_1\lesssim1.6$ ($\PR=0.015$--0.030 at $N=4000$), weakens at $k_c=1.8k_1$ ($\PR=0.07$), and is absent at $k_c=2k_1$, where only an ordinary minimum remains near $q^*$ and the mode just below it stays extended ($\PR\simeq0.2$ for $N=500$--4000; main-text Fig.~4). At $N=500$, the quartic resonance lies at $3.555\,\omega_0$, just below $\omega^*$, with $\PR=0.051$ and $W_b=0.66$; its four decaying factors have $\xi/a\simeq9$, and the effective participating volume $2N\PR$ stays near 50 sites for $N=500$--1000 and then grows to 88 at $N=4000$, rather than saturating as for a bound state.

The channel count does not by itself decide whether a resonance forms: $d=2$ near both the minimal-ladder saddle-node and self-pairing, and $d=4$ near both the cubic and the quartic point of the longer-range ladder. What differs is whether the decaying factors have a propagating partner in their own cluster. Odd-multiplicity (cubic) roots always supply one, which locks the decaying amplitude and depletes the boundary; even-multiplicity (quartic) roots do not, and produce a resonance whose $\PR$ keeps decreasing with $N$ over the accessible range.

\section{Experimental specimen}
\label{sec:exp}

The specimen is a three-dimensional phononic structure that reproduces the essential ingredients of the ladder model, namely nearest- and next-nearest-neighbor couplings on chain A and only nearest-neighbor coupling on chain B [main-text Fig.~5(a),(b)]. It is composed of alternating blocks and frames, which serve as the discrete masses of the system; each block is attached to a surrounding frame through structural bridges. The block width is $W_m=8.06$~mm, and the square frame width is $W_f=19.35$~mm.

Navy rods ($R_1=3.23$~mm, $L_1=6.45$~mm) implement $k_1$ on both chains, inclined green rods ($R_2=3.87$~mm) implement $k_2$ on chain A, a pink bridge ($L_c=15$~mm) implements $k_c$, and axial orange bridges ($L_w=50$~mm) connecting each chain-B mass to the rigid outer frame implement $k_w=2k_2$ on chain B, providing the effective grounding stiffness that matches the diagonal contribution of the NNN coupling on chain A [main-text Fig.~5(b)]. At the terminal cells, extended NNN bridges on chain A and two long struts connecting sites $1A$ and $1B$ to the outer frame restore the diagonal stiffness lost with the cut bonds. The equivalent stiffness of the struts is not independently calibrated, so the termination is bulk matched only approximately; given the sensitivity found in Sec.~\ref{sec:boundary}, we analyze only the lower end. Continuum simulations use COMSOL with Young's modulus 1.8~GPa and density 1300~kg/m$^3$. The specimen, clamped at its outer frame, was driven at the lower terminal cell by an electrodynamic shaker (PCB 2007E01, Krohn-Hite 7500 amplifier), and the velocity of both chains in all 20 cells was measured by SLDV. The dispersion follows from a two-dimensional space--time Fourier transform of the velocity difference between the two ends of each frame.

The specimen realizes the ladder through two torsional branches of the frames [main-text Fig.~5(c),(d)]; the dashed curves in main-text Fig.~5(c) are other deformation modes, which the SLDV cannot access in this setup. Because each frame is torsionally restrained by the outer frame, both branches start at nonzero frequency, and their proportions differ from the model: the computed upper branch peaks at 1482~Hz near $qa\approx1.66$, about 25\% above $f_\pi=1181.5$~Hz, against 2\% in the model at $\gsp$. The correspondence is therefore in band shape---avoided crossing, interior maximum, and flat zone boundary---so we place the specimen with dimensionless band-shape quantities that use no reference frequency.

Over $\pi-0.38\leq qa\leq\pi$ the computed upper branch varies by only 1.4~Hz. With $1/\meff=d^2\omega/dq^2=2\pi\,d^2f/dq^2$, polynomial fits near $qa=\pi$ give $d^2f/dq^2=-48\pm6$~Hz\,$a^2$, against $-(1000$--$1250)$~Hz\,$a^2$ at the interior maximum, so the inverse-effective-mass ratio is $\rho\equiv|1/\meff(\pi/a)|/|1/\meff(q_{\max})|=0.038$--0.048. The negative sign makes the zone boundary a weak local maximum, with an interior minimum at $q_sa\approx2.84$, about 1.4~Hz below $f_\pi$: the configuration of main-text Fig.~1(b) for $g<\gsp$. The specimen thus satisfies main-text Eq.~(2) with $k=1$ exactly at $\pi/a$ and with $k=2$ to within 5\% of the inverse effective mass at the interior maximum. Its separation $\pi-q_sa=0.30$ and relative depth $\delta\equiv[f_\pi-f(q_s)]/(f_{\max}-f_\pi)=0.0049$ give two further measures. In the model at $\eta=1$, $(\rho,\,\pi-q_sa,\,\delta)=(0.058,\,0.33,\,0.021)$, $(0.038,\,0.27,\,0.0089)$, and $(0.019,\,0.19,\,0.0021)$ at $g/\gsp=0.85$, 0.9, and 0.95. Matching each quantity separately gives $g/\gsp=0.87$--0.90 from $\rho$, 0.88 from $\pi-q_sa$, and 0.93 from $\delta$, so we place the specimen at $g/\gsp\approx0.9$, where the minimal-ladder resonance retains the same channel content as the resonance at critical coupling (Sec.~\ref{sec:finite}).

The boundary-weighted response at 1156~Hz lies about 25~Hz below $f_\pi$, on the side where the nearly coalesced roots near $c=-1$ are complex, and well below the interior maximum, so it is embedded in the continuum. This is the regime the experiment is designed to probe. Because the weakly evanescent channels and the resulting resonance persist slightly below criticality (Secs.~\ref{sec:roots} and \ref{sec:finite}), the specimen need not sit exactly at the quartic point; it must realize a nearly flat zone boundary close to $\gsp$, which the band-shape analysis above establishes. That analysis uses the computed branch, which the measured dispersion reproduces, because a variation of 1.4~Hz over the wavenumber spacing $2\pi/N\approx0.31$~rad of a 20-cell specimen lies below the damping width $2\zeta f_\pi\approx24$--71~Hz of PLA ($\zeta\approx0.01$--0.03). The measured profiles then test the central prediction directly: boundary enhancement appears just below the flat zone boundary and is absent away from it.

The actuated cell $n=1$ is included in the dispersion and in the displayed profiles of main-text Fig.~5(f),(h) but excluded from the boundary weight, because its motion is imposed by the drive. We define
\begin{equation}
	W_b^{\mathrm{low}}=\frac{\sum_{n=2}^{3}(|u_{nA}|^2+|u_{nB}|^2)}{\sum_{n=2}^{20}(|u_{nA}|^2+|u_{nB}|^2)},
	\label{eq:wlow}
\end{equation}
whose uniform value is $2/19\approx0.11$. We find $W_b^{\mathrm{low}}=0.59$ at 1156~Hz and 0.25 at 1000~Hz. Part of the weight at either frequency may come from proximity to the drive, which plausibly accounts for the moderate enhancement at 1000~Hz; because both responses are driven at the same cell with the same configuration, proximity cannot account for the further factor of 2.4 between them.

\end{document}